# Measuring sheet resistance of CIGS solar cell's window layer by spatially resolved electroluminescence imaging

Myriam Paire, Laurent Lombez, Jean-François Guillemoles, Daniel Lincot,

*Institute of research and development for photovoltaic energy, UMR 7174 CNRS/EDF R&D/Chimie ParisTech, 6 quai Watier, 78401 Chatou, France*

Correspondence should be addressed to myriam-paire@chimie-paristech.fr or laurent-lombez@chimie-paristech.fr

**Abstract :**

A spatially resolved electroluminescence (EL) imaging experiment is developed to measure the local sheet resistance of the window layer, directly on the completed CIGS cell. Our method can be applied to the EL imaging studies that are made in fundamental studies as well as in process inspection[1-4]. The EL experiment consists in using solar cell as a light emitting device : a voltage is applied to the cell and its luminescence is detected. We develop an analytical and quantitative model to simulate the behavior of CIGS solar cells based on the spread sheet resistance effect in the window layer. We determine the repartition of the electric potential on the ZnO, for given cell's characteristics such as sheet resistance and contact geometries. Knowing the repartition of the potential, the EL intensity is measured and fitted against the model. The procedure allows the determination of the window layer sheet resistance.

**Introduction :**

The EL spatially resolved imaging is a convenient characterization tool to analyse solar cells or modules. Camera-based EL experiments are widely used as on-line quality tests in the industry, but are also used in laboratories to gain more insight in loss mechanisms[1-3, 5, 6]. As it was previously suggested[1, 4], it is indeed possible to extract the value of the window layer sheet resistance from an electroluminescence image.

The present paper focuses on the determination of the window layer sheet resistance in CIGS solar cells, but our approach could be transferred to other photovoltaic technologies, especially to thin films where the determination of sheet resistance is difficult on the completed device. An appropriate model of the sheet resistance effect, in terms of voltage drop on the cell's surface, is developed. From an EL spatially resolved image, the decay of the EL intensity with respect to the distance from the electrode is determined and fitted using our model. A value of the sheet resistance on the completed cell is therefore accessible.

**Experiment**

We conduct an EL experiment on a CIGS solar cell. The structure of the cell is Soda-lime glass/Mo/CIGS/CdS/i:ZnO/ZnO:Al, where the ZnO layers are deposited by sputtering. The EL experiment consists in using the cell as a light emitting device. We use a two points probe configuration and a Keithley 2635A source meter. A tungsten probe is placed on the Mo back contact, the other one is centred on the cell. Under dark conditions, we apply a voltage on the cell. Once the applied voltage is sufficient, we detect a luminescence signal. The intensity of the luminescence is recorded by a CCD camera, which can provide a 2D spatially resolved image of the luminescence of the cell. We integrate the signal on an angular sector in order to account for material inhomogeneities and get an average value of the luminescence at a certain distance of the probe. We remark that the luminescence is decreasing with respect to the distance from the front contact probe. We propose to evaluate quantitatively the sheet resistance of the window layer from the variation of the luminescence signal.

**The model**

The luminescence intensity $\Phi$ is related to the voltage V that is applied between the back and front contact of a cell:

$$\phi = \phi_0 \exp(qV/kT) \tag{1.}$$

where kT/q is the thermal voltage and $\Phi_0$ a calibration factor that depends on, among others factors, the camera calibration settings, the device optics, the diffusion lengths and surface recombination velocities. Therefore when the front contact sheet resistance cannot be neglected, the voltage on the cell surface is a function of the distance from the contact V=V(r) and, as a consequence, so is the luminescence $\Phi = \Phi(r)$. In order to quantify this sheet resistance effect from the electroluminescence experiment, we develop an analytical model.

The detailed description of the model is given elsewhere[7]. We limit our study at the window layer of the cell, which is supposed to be resistive with a certain sheet resistance $R_\square$. We suppose that the cell characteristics are well described by a one-diode model. Therefore the current density coming from the p-n junction is:

$$Jz(z=0) = -J_0(\exp(q\,\psi(z=0)/nkT) - 1) - \psi(z=0)/R_{sh} \qquad (2.)$$

where $\psi$ is the electric potential in the window layer and z the altitude, which is set to 0 at the bottom of the window layer and equals t at the surface, $J_0$, n, kT/q, $R_{sh}$ are respectively the diode saturation current density, the diode ideality factor, the thermal voltage, and the shunt resistance. The series resistance of the junction itself is neglected as its effect is smaller than that of the sheet resistance. In order to get the values of those parameters for the cell studied, we apply a simple one-diode fit on a dark current-voltage measurements that is made in-situ before the beginning of the electroluminescence experiment.

Our goal is to get the repartition of the potential on the cell surface with respect to the distance from the electrical contacts. For the sake of simplicity, we solve the problem in a cylindrical symmetric situation, where the probe contact is a tip, placed at the center of a circular cell, but the results can easily be extended to other geometries such as parallel grid lines. We show that the potential $\psi$ on the surface of the cell is the solution of the following one-dimensional equation[7]:

$$\partial^2\psi/\partial r^2 + 1/r \times \partial\psi/\partial r - R_\square(J_0(\exp(q\psi/nkT)+1)) - \psi/R_{sh} = 0 \qquad (3.)$$

where r is the radial distance from the probe tip, $R_\square = \rho/t$ is the sheet resistance (Ohm) and $\rho$ the resistivity of the window layer. This differential equation can easily be solved by a standard solver software. The boundary conditions that we used to solve equation (3.) were the following. As the probe tip is an equipotential at the applied voltage V, $\psi(b) = V$ where b is the probe tip radius. In the absence of recombination current at the cell perimeter, we have the Neumann condition $\partial\psi/\partial r(a) = 0$, where a is the cell's radius.

For a given sheet resistance $R_\square$ of the window layer, the potential repartition on the cell front surface is determined by solving equation (3.), and the corresponding electroluminescence intensity is deduced from equation (1.). Note that our method differs from the one proposed in [4], where the potential is deduced from the electroluminescence signal.
 In order to compare the luminescence signal from our simulated data with the experimental ones, we normalize both luminescence signals by their maximum value. We fit the experimental data with our model by adjusting the value of $R_\square$, which is the single fit parameter. Therefore we extract from the luminescence decay the value of the sheet resistance of the window layer of the cell under study.
.

**Results**

We study CIGS solar cells with the classical structure Soda-lime glass/Mo/CIGS/CdS/i:ZnO/ZnO:Al, where the ZnO:Al layer is 400 nm thick. We analyze square solar cells which were mechanically scribed from the same larger substrate of two different areas : 0,1 and 0,3 cm². We simulate these cells by circular cells of the same area. We record the EL image for different applied voltages. We fit the EL intensity signal by our model and give the results below.

In order to fit the EL signal we need to know the effective applied voltage at the ZnO surface. Therefore we need a tool to determine the contact resistance between the probe tip and the ZnO surface, as this resistance was found very important in our experimental setup. We record several EL images at different reference applied voltages. Then we fit these signals in order to extract two parameters, the effective applied voltage and the sheet resistance. We validate our model by verifying that the difference between the reference and effective applied voltages is due to a contact resistance that we evaluate.

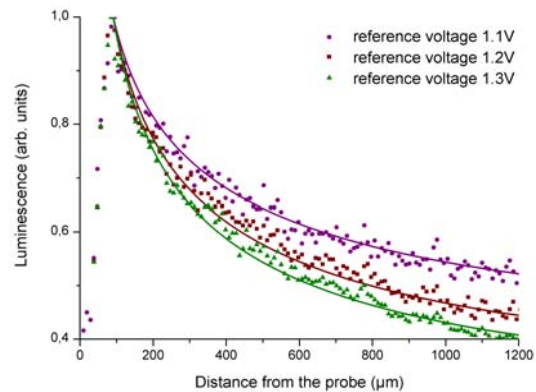

**Figure 1 : EL data (dots) and fit (solid line) at three different reference voltages (1.1V, 1.2V, 1.3V) applied to a 0.3cm² solar cell**

Figure 1 shows the results of the EL experiment at three different reference voltages for a 0.3 cm² solar cell (1.1V, 1.2V and 1.3V). For the three experiments, the fits give the same value of the sheet resistance, that is 30 Ohm/square +/- 1 Ohm and the corresponding "effective" applied voltages. The latter are the reference applied voltages corrected by the voltage drops due to the contact resistance effect. The contact resistance taking place at the interface between the probe's tip and the ZnO surface, is obtained from the effective and reference voltages by :

$R_{contact} = (V_{ref}-V_{eff})/I(V_{ref})$ (4.)

where $R_{contact}$ is the contact resistance, $V_{ref}$ the reference voltage, $V_{eff}$ the effective voltage, and $I(V_{ref})$ the intensity measured on the in-situ dark I-V measurement at the reference applied voltage $V_{ref}$.

| Reference voltage (V) | Effective voltage (V) | Intensity at the reference voltage (mA) | Contact resistance (Ohm) |
|---|---|---|---|
| 1.1 +/- 0.1% | 0.72 +/- 5% | 4,81 +/- 0.1 % | 79 +/- 5% |
| 1.2 +/- 0.1% | 0.74 +/- 5% | 5,89 +/- 0.1% | 78+/- 5% |
| 1.3 +/- 0.1% | 0.75 +/- 5% | 6,94 +/- 0.1% | 79 +/- 5% |

**Table 1 : Reference voltage, effective applied voltage, intensity at the reference voltage and the calculated contact resistance from the three EL images on the 0.3 cm² solar cell.**

One can see that the difference between the reference voltage and the effective voltage is due to a contact resistance effect. The contact resistance, in our contacting geometry with a tungsten probe is approximately 79 Ohm. Therefore it is an important effect that has to be taken into account.

The fit of our EL images gives us access to the effective voltage applied on the ZnO surface but more importantly to the sheet resistance of the window layer of the cell. We studied cells of various sizes to test the validity of our model. We give the results found for two different cells (0.1cm² and 0.3cm²).

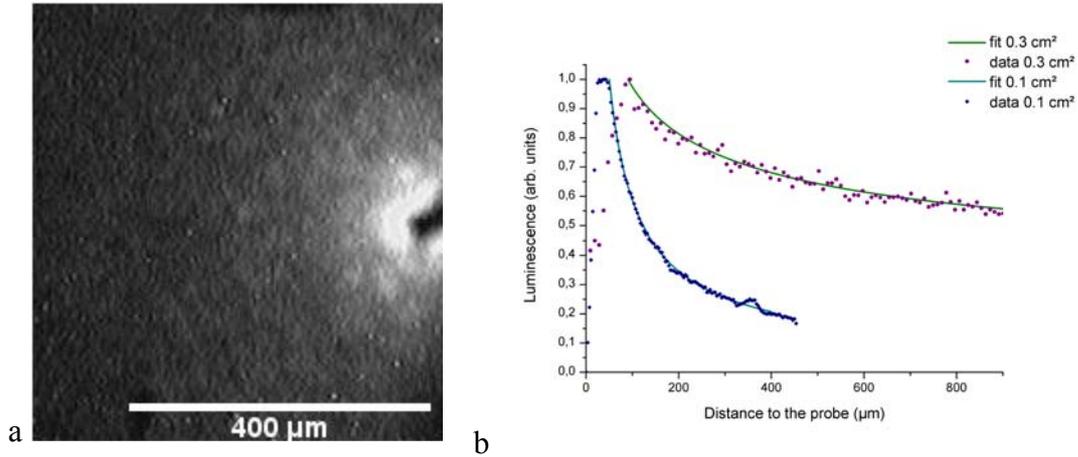

a  b

**Figure 2 : a- EL spatially resolved image of the 0.1cm² cell under 0.76V. b- Electroluminescence experimental (dots) and simulated (solid line) signals as a function of the distance from the probe. The circular dots corresponds to the 0.3 cm² cell under an applied voltage of 0.72V, the diamond-shapes to the 0.1 cm² cell under an applied voltage of 0.76 V.**

On figure 1 and 2, one can notice that for distances from the probe tip below 50 µm, the electroluminescence signal is dimmed. This is due to the fact that the probe creates a shadow which attenuates the signal. The data used to simulate the EL signal are extracted from a current-voltage measurement done prior to the EL experiment. From a simple one-diode fit, we found for example for the 0.1 cm² cell $I0 = 2.974 \cdot 10^{-9}$

A, n = 1,815. As the shunt resistance is $2 \cdot 10^5$ Ohm, we decided to neglect its influence and we consider an infinite shunt resistance. The best fit is obtained for a sheet resistance of 30 Ohm/square +/- 1 Ohm/square for the different voltages (0.71V, 0.73V, 0.76V) that we applied on this 0.1 cm² solar cell. Therefore our model is able to take into account variations in the EL intensity due to different applied voltages. We proceed in the same manner for the 0.3 cm² solar cell. The data extracted from the current-voltage measurement are $I0 = 5.137 \cdot 10^{-10}$ A, n = 1,718. We fit the EL luminescence signal at three applied voltage (0.72V, 0.74V, 0.75V) and found that the sheet resistance is again 30 Ohm/square +/- 1 Ohm/square. This is coherent with the results of the 0.1 cm² cell as the two cells were mechanically scribed on the same substrate, and therefore have the same window layer.

This sheet resistance value of 30 Ohm/square +/- 1 Ohm/square is also coherent with the one we found with a classic current-voltage experiment under AM1.5 illumination. We use a solar simulator to illuminate the cell, and we measure a current-voltage curve. We adapt our model for the interpretation of experiment under illumination by introducing a photocurrent density term. The detailed equations of our model for the illuminated case are given elsewhere[7]. We are able to extract from the current-voltage curve the value of the sheet resistance, which is 34 Ohm/square +/- 2 Ohm/square (see figure 3) for the 0,1 cm² cell. The coherence between the EL experiment and the illuminated current-voltage measurement confirms that the value of the sheet resistance extracted from the EL is valid and accounts for the sheet resistance effect visible on well-known current-voltage curve. The difference between the two experiments can originate from the two different experimental setups, as well as the photocurrent not being strictly constant with the voltage, as supposed in our model.

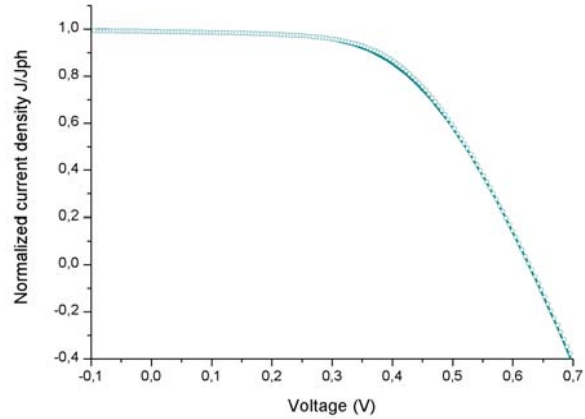

**Figure 3 : Current-voltage under AM1.5 for the 0.1 cm² cell. Dots are the experimental data, the line is our model's fit.**

We then compare those values of window layer sheet resistance with the one available from a ZnO:Al layer deposited on glass under the same conditions. With a four point probe experiment on a ZnO:Al layer of 420nm thickness on glass we found that the sheet resistance is 14,7 Ohm/square +/- 0,3 Ohm/square. The cells we study in the EL experiment have a ZnO:Al layer that is deposited in the same conditions, in terms of time and sputtering characteristics. Therefore it appears that the data of our fit gives a value of the sheet resistance that is higher than the one measured from ZnO:Al on glass. This can stem from different considerations. First, the ZnO:Al layer that is grown on the cell differs from the one grown on glass in terms on material properties, as the nucleation on ZnO:i and on glass is different resulting in different material properties, such as crystallites sizes and grain boundary properties for example.

Therefore the lateral conductivity of the ZnO:Al should differ in the two layers. Second, the cell is a rougher substrate than the glass. If the surface is not flat the average distance to the electrode is increased by the surface roughness. This can lead to an apparent sheet resistance that is more important than the one of a flat cell of the same area. Therefore the apparent sheet resistance of the window layer on the cell is higher than the one of the same layer on a flat glass surface. Third, in our model we neglect the internal series resistance of the solar cell. If this resistance is not negligible compared to the sheet resistance, it can artificially increase our estimated sheet resistance. Therefore more experiments, with window layers of different sheet resistance, will be conducted in order to determine the reasons for the difference between the sheet resistance of a window layer on the cell and on glass.

**Conclusion**

We developed a model that is able to estimate the voltage at the cell surface once given the sheet resistance of the window layer, the cell's diode characteristics and the applied voltage. Therefore we are able to interpret the decay of EL signal with the distance from the electrode as a sheet resistance effect, and we can give a quantitative estimation of the sheet resistance value. If this work has been done considering circular cells for the sake of simplicity, a development towards other geometries such as parallel grid lines can easily be made. Further studies will also concentrate on new contacting methods where the contact resistance will be accurately known and on the study of cells covered with window layers of different properties in order to gain more insight in the link between the properties of the window layer deposited on glass and on the cell. In addition to the estimation of series and shunt, it is thought that EL imaging can be developed to estimate the sheet resistance for completed cells or modules.

**References**

1. Helbig, A., Kirchartz, T., Schäffler, R., Werner, J.H., Rau, U., 24th European Photovoltaic Solar Energy Conference, Hamburg, Germany, 2009: p. 2446-2449.
2. Sugimoto, H., Aramoto, T., Kawaguchi, Y., Chiba, Y., Kijima, S. Fujiwara, Y., Tanaka, Y., Hakuma, H., Kakegawa, K., Kushiya, K., 24th European Photovoltaic Solar Energy Conference, Hamburg, Germany, 2009: p. 2465-2468.
3. Weber, T., Kutzer, M., 24th European Photovoltaic Solar Energy Conference, 21-25 September 2009, Hamburg, Germany, 2009: p. 477-479.
4. Helbig, A., et al., Solar Energy Materials and Solar Cells, 2010. **94**(6): p. 979-984.
5. Kirchartz, T., et al., Thin Solid Films, 2007. **515**(15): p. 6238-6242.
6. Fuyuki, T., et al., Applied Physics Letters, 2005. **86**(26): p. 262108.
7. Paire, M., Lombez, L., Guillemoles, J.F., Lincot, D., Journal of Applied Physics, 2010 - in print.